\documentclass[traditabstract,longauth]{aa}

\usepackage{graphicx}
\usepackage{txfonts}
\usepackage{natbib}
\newcommand{\um}{$\mu$m}

\begin{document}
   \title{A \emph{Herschel} study of YSO evolutionary stages and formation timelines in two fields of the Hi-GAL survey \thanks{\emph{Herschel} is an ESA space observatory with science instruments provided by European-led Principal Investigator consortia and with important participation from NASA.}}

   \subtitle{}

   \author{D. Elia\inst{\ref{ifsi},\ref{lisboa}}
          \and
          E. Schisano\inst{\ref{ifsi},\ref{nap}}
          \and
          S. Molinari\inst{\ref{ifsi}}
          \and
          T. Robitaille\inst{\ref{harv}}
          \and
          D. Angl\'{e}s-Alc\'{a}zar\inst{\ref{ariz}}
          \and
          J. Bally\inst{\ref{Color}}
          \and
          C. Battersby\inst{\ref{Color}}
          \and
          M. Benedettini\inst{\ref{ifsi}}
          \and
          N. Billot\inst{\ref{Pasadena}}
          \and
          L. Calzoletti\inst{\ref{asi}}
          \and
          A. M. Di Giorgio\inst{\ref{ifsi}}
          \and
          F. Faustini\inst{\ref{asi}}
          \and
          J. Z. Li\inst{\ref{pechino}}
          \and
          P. Martin\inst{\ref{canada1},\ref{canada2}}
          \and
          L. Morgan\inst{\ref{liv}}
          \and
          F. Motte\inst{\ref{saclay}}
          \and
          J. C. Mottram\inst{\ref{exeter}}
          \and
          P. Natoli\inst{\ref{rm2}}
          \and
          L. Olmi\inst{\ref{Prico},\ref{arcetri}}
          \and
          R. Paladini\inst{\ref{ipac}}
          \and
          F. Piacentini\inst{\ref{rm1}}
          \and
          M. Pestalozzi\inst{\ref{ifsi}}
          \and
          S. Pezzuto\inst{\ref{ifsi}}
          \and
          D. Polychroni\inst{\ref{ifsi}}
          \and
          M. D. Smith\inst{\ref{kent}}
          \and
          F. Strafella\inst{\ref{lecce}}
          \and
          G. S. Stringfellow\inst{\ref{Color}}
          \and
          L. Testi\inst{\ref{eso}}
          \and
          M. A. Thompson\inst{\ref{hfs}}
          \and
          A. Traficante\inst{\ref{rm2}}
          \and
          M. Veneziani\inst{\ref{rm1}}
          }

   \institute{Istituto di Fisica dello Spazio Interplanetario - INAF,  via Fosso del Cavaliere 100, 00133 Rome\label{ifsi}\\
              \email{davide.elia@ifsi-roma.inaf.it}
         \and
             Observat\'orio Astron\'omico de Lisboa, Tapada da Ajuda, 1349-018 Lisboa, Portugal\label{lisboa}
         \and
             Dipartimento di Scienze Fisiche, Universit\`{a} di Napoli ``Federico II'', Naples, Italy\label{nap}
         \and
             Harvard-Smithsonian Center for Astrophysics, 60 Garden Street, Cambridge, MA 02138, USA\label{harv}
         \and
             Department of Physics, University of Arizona, 1118 E. 4th Street, Tucson, AZ 85721 \label{ariz}
         \and
             University of Puerto Rico, Rio Piedras Campus, Physics Dept., Box 23343, UPR station, San Juan, Puerto Rico\label{Prico}
         \and
             Center for Astrophysics and Space Astronomy, University of Colorado, Boulder, CO 80309-0389, USA\label{Color}
         \and
             NASA \emph{Herschel} Science Center, IPAC, Caltech, Pasadena, CA 91125, USA\label{Pasadena}
         \and
             ASI Science Data Center, I-00044 Frascati (Rome), Italy \label{asi}
         \and
             National Astronomical Observatories, Chinese Academy of Sciences, Beijing 100012, China \label{pechino}
         \and
              Canadian Institute for Theoretical Astrophysics, University of Toronto, 60 St. George Street, Toronto, ON M5S 3H8,
              Canada\label{canada1}
         \and
              Department of Astronomy \& Astrophysics, University of Toronto, 50 St. George Street, Toronto, ON M5S 3H4, Canada\label{canada2}
         \and
              Astrophysics Research Institute, Liverpool John Moores University, Twelve Quays House, Egerton Wharf, Birkenhead CH41 1LD\label{liv}
         \and
              Laboratoire AIM, CEA/IRFU – CNRS – Universit\'{e} Paris Diderot, Service d'Astrophysique, 91191 Gif-sur-Yvette, France\label{saclay}
         \and
              School of Physics, University of Exeter, Stocker Road, Exeter, EX4 4QL, United Kingdom\label{exeter}
         \and
              INAF, Osservatorio Astrofisico di Arcetri, Largo E. Fermi 5, I-50125, Firenze, Italy\label{arcetri}
         \and
              Spitzer Science Center, IPAC, MS 220-6, California Institute of Technology, Pasadena, CA 91125, USA\label{ipac}
         \and
              Centre for Astrophysics and Planetary Science, University of Kent, Canterbury CT2 7NH, UK\label{kent}
         \and
              Dipartimento di Fisica, Universit\`{a} del Salento, CP 193, I-73100 Lecce, Italy\label{lecce}
         \and
             ESO, Karl Schwarzschild-Strasse 2, 85748 Garching bei M\"{u}nchen, Germany\label{eso}
         \and
             Centre for Astrophysics Research, University of Hertfordshire, College Lane, Hatfield AL10 9AB, UK\label{hfs}
         \and
             Dipartimento di Fisica, Universit\`{a} di Roma 2 ``Tor Vergata'', Rome, Italy\label{rm2}
         \and
             Dipartimento di Fisica, Universit\`{a} di Roma 1 ``La Sapienza'', Rome, Italy   \label{rm1}
             }

\date{}

  \abstract{We present a first study of the star-forming compact dust
condensations revealed by \emph{Herschel} in the two $2 \degr \times 2
\degr$ Galactic Plane fields centered at $[\ell,b]=[30^{\circ},0^{\circ}]$ and
$[\ell,b]=[59^{\circ},0^{\circ}]$, respectively, and observed during the Science
Demonstration Phase for the Herschel infrared Galactic Plane survey (Hi-GAL) Key-Project.
Compact source catalogs extracted for the two fields in the five Hi-GAL
bands (70, 160, 250, 350 and 500\um) were merged based on simple criteria
of positional association and spectral energy distribution (SED) consistency
into a final catalog which contains only coherent SEDs with counterparts
in at least three adjacent Herschel bands. These final source lists contain
528 entries for the $\ell=30^{\circ}$ field, and 444 entries for the
$\ell=59^{\circ}$ field. The SED coverage has been augmented with ancillary
data at 24~\um\ and 1.1~mm. SED modeling for the subset of 318 and 101 sources
(in the two fields, respectively) for which the distance is known was
carried out using both a structured star/disk/envelope radiative transfer
model and a simple isothermal grey-body. Global parameters like mass,
luminosity, temperature and dust properties have been estimated. The
$L_{\textrm{bol}}/M_{\textrm{env}}$ ratio spans four orders of magnitudes
from values compatible with the pre-protostellar phase to embedded massive
zero-age main sequence stars.
Sources in the $\ell=59^{\circ}$ field have on average lower $L/M$, possibly
outlining an overall earlier evolutionary stage with respect to the sources
in the $\ell=30^{\circ}$ field. Many of these cores are actively forming
high-mass stars, although the estimated core surface densities appear to be
an order of magnitude below the 1~g~cm$^{-2}$ critical threshold for
high-mass star formation.}


   \maketitle
%

\section{Introduction}
The chance to observe the far-infrared emission from cold dust condensations in star-forming
regions is one of the most important improvements offered by the PACS \citep{pog10} and SPIRE
\citep{gri10} cameras on board the ESA \emph{Herschel} Space Observatory \citep{pil10}. The \emph{Herschel}
Infrared Galactic Plane Survey \citep[Hi-GAL][]{mol10a,mol10b} is a \emph{Herschel} Key-Project that maps
the inner Galaxy ($| \ell |< 60^{\circ}, | b |< 1^{\circ}$) in five photometric bands
(centered at 70, 160, 250, 350, and 500~$\mu$m) and has been designed mainly for observing star-forming
regions and cold interstellar medium (ISM) structures with unprecedented spatial resolution. Thanks to its capabilities
in terms of spectral coverage, resolution and sensitivity, it is expected to provide a huge progress in
the study of the early stages of star formation, and in particular for disclosing new knowledge about the
formation of massive stars.

Indeed, the mass of a young stellar object (YSO) plays a critical role in distinguishing between various
possible regimes and the timescales of evolutionary paths. Whereas the sequence for the formation of
Solar mass objects is quite well-defined, for high-mass protostars ($M \ge 8~M_{\odot}$) it is not completely
clear yet, due to the rarity of these sources in the Solar neighborhood ($d\leq 1$ kpc) and their shorter
evolutionary timescales (it is expected that such stars begin burning hydrogen while still
accreting mass from the parental gas envelope).

The analysis of the infrared spectral energy distribution (SED) is a powerful method used to classify
YSOs from the evolutionary point of view, and indeed the classification scheme (coded in Classes
from 0 to~III) proposed by \citet{lad84} and refined by \citet{and93} is now well-established and commonly
used for characterizing the population of star-forming regions. \citet{sar96} derived physical
properties from SED fits, and plotted in a diagnostic diagram the bolometric luminosity vs the envelope mass
to easily categorize the YSOs and describe the entire Class~0-II evolution. More recently \citet{mol08}
extended this analysis to massive YSOs, and demonstrated that it is applicable and meaningful also
in this case. In this paper, we intend to apply the same methodology to the source samples extracted
from the Hi-GAL maps obtained during the \emph{Herschel} Science Demonstration Phase (SDP), and
to perform a very first test on its ability to describe the star formation timeline of the
investigated regions. We describe how the SEDs were assembled and how models were fitted to data
to estimate the physical parameters of the sources. Finally, we briefly discuss
the $L_{\rm bol}/M_{\rm env}$~vs~$M_{\rm env}$ diagram.

\section{SED building}\label{sedb}
For each of the two SDP fields, a first catalog based on the Hi-GAL image photometry has been compiled, identifying the sources detected in the five different bands \citep[see][for details]{mol10b} based on simple positional association.
Starting from the source extraction list at the longest wavelength, namely 500~$\mu$m, an association with the previous band (in this case, 350~$\mu$m) was established if a source was found within a search radius corresponding to the \emph{Herschel} half-power beam-width at the longer of the two considered wavelengths; if multiple associations were found, the closest source was kept. Other counterparts fulfilling the positional association criterium were not merged, but were added anyway as independent sources and considered as such in the following steps.
The celestial coordinates assigned to the merged sources are those corresponding to the shorter wavelength association, which is by definition characterized by higher spatial resolution. After having correlated the first two wavelengths, a list composed by associated sources (hereafter considered as the same source) and by unassociated single-band detections in both wavelengths was obtained. Entries in this list were then correlated with the next wavelength source list according to the same criteria, and so on, down to 70~$\mu$m. The occurrences of multiple associations tend to be more frequent with decreasing wavelength (and increasing resolution), but for the time being we did not attempt to split up the flux at any given band into the contributions by different counterparts detected at the shorter wavelengths.

The catalogs resulting at the end of this procedure (which we will call \textsl{Stage 1} catalogs) contain 2022 entries for the $\ell=30^{\circ}$ field, and 1322 entries for the $\ell=59^{\circ}$ field. Entries in these catalogs are in principle SEDs exhibiting a variable number of counterparts for the various Herschel bands for which the positional association was successful. An investigation of the resulting SEDs reveals a variety of situations in which the positional association clearly did not work properly; this is apparent in SEDs which have missing bands and/or deep concavities due to a flux that is inconsistent with adjacent ones. This may happen either because the simple positional association failed, or because the flux extracted for that source in the detection and photometry stage \citet{mol10b} in that particular band was corrupted (e.g. the source fitting did not converge because of source crowding or particularly difficult background conditions). These irregular SEDs were excluded from further analysis, as they are unreliable. The remaining sources populate the so-called \textsl{Stage 2} catalog (see Table~\ref{stats}). The subsequent selection step is two-fold and generates on the one hand a \textsl{Stage 3a} catalog containing sources detected in at least the four 70,160, 250 and 350 \um\ bands, which will be used for the color-color analysis. On the other hand, for the purpose of SED modeling (see Sect.~\ref{sedfit}) we choose sources which have at least three bands, but for which a distance is available \citep{rus10}; this will generate the \textsl{Stage 3b} catalogs.

\begin{table}
\caption{Source catalog statistics}
\label{stats}
\centering
\begin{tabular}{l | l c c}
\hline\hline
 & & $\ell=30^{\circ}$ & $\ell=59^{\circ}$ \\
\hline
\textsl{Stage 1} & \parbox{5cm}{All sources detected in at least one PACS/SPIRE band} & 2022 & 1322  \\
\textsl{Stage 2} & \parbox{5cm}{Stage 1 sources excluding those with irregular SEDs} & 528    & 444 \\
\textsl{Stage 3a} & \parbox{5cm}{Stage 2 sources identified in at least the 70, 160, 250 and 350\um\ bands} & 236  & 141 \\
\textsl{Stage 3b} & \parbox{5cm}{Stage 2 sources identified in at least three bands, but also with a known distance} & 318 & 101 \\
\hline
\end{tabular}
\end{table}

\begin{figure}
   \centering
   \includegraphics[width=6.3cm]{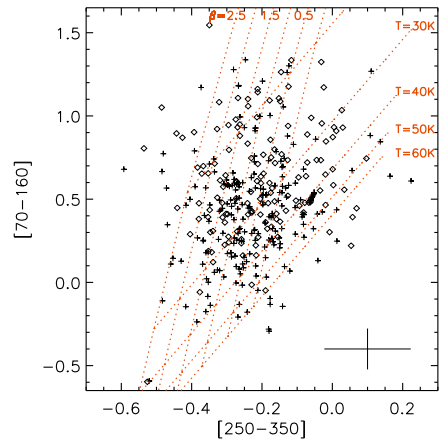}
   \caption{Two-color diagram [70--160] vs [250--350] for the \textsl{Stage 3a} sources cataloged in the two investigated fields (crosses for $\ell=30^{\circ}$, and diamonds for $\ell=59^{\circ}$, respectively).
   ~
   Red dotted lines represent the loci of greybodies for various constant values of temperature and emissivity law exponent. In the bottom right corner the average error bar is reported.}
    \label{colcol}
    \end{figure}

In Fig.~\ref{colcol} a [70--160] vs [250--350] two-color diagram is presented for the \textsl{Stage 3a} catalog. For comparison, a grid representing the isothermal greybody model loci is also overplotted. The analytical expression of a greybody is

\begin{equation}
S_{\nu}=\frac{M~\kappa_0}{d^{2}}\left(\frac{\nu}{\nu_0}\right)^{\beta}B_{\nu}(T) \;,
\label{greyb}
\end{equation}
where $M$ is the total (gas + dust) core mass, $\kappa_{0}$ is the total mass absorption coefficient evaluated at a fixed frequency $\nu_0$ \citep[in this case $\kappa_{0} = 0.005$~cm$^{2}$~g$^{-1}$ at $\nu_0=230$~GHz,][]{pre93}, $d$ is the distance to the object, $\beta$ is the dust emissivity index, and $B_{\nu}(T)$ is the Planck function.

Although there is no evidence of segregation, we can affirm that the colors are consistent with greybodies having temperatures in the range of $20 \lesssim T \lesssim 60$~K. We note however that a noticeable fraction of sources lies in the region of the plot corresponding to negative $\beta$ values. One possibility is that the single temperature assumption in Eq.~\ref{greyb} is incorrect and that a multicomponent fit would be warranted. This is in line with the accepted scenario in which envelopes around YSOs show a radial temperature gradient. The other possibility is that we are overestimating the flux at longer wavelengths (thus flattening the SED in the submillimeter range and mimicking a low or negative $\beta$) because we are associating counterparts whose size is increasing at increasing wavelengths. Both effects are probably playing a role here.

An alternative approach in building SEDs would have been to use the same size at all wavelengths, but this would have meant either degrading the map resolution to a common grid (the 500\um\ one), or forcing the detection at 350 and 500\um\ using the sources detected shortward of 250\um\ and constraining the flux estimates to the same source size as measured at 250\um\ (and taking the different beam sizes into account). There are disadvantages in both cases; besides, the flux extraction from the 350 and 500\um\ maps on the locations of the 250\um\ detected sources is not at all straightforward. These alternative possibilities will be investigated in subsequent works.



\section{SED fitting}\label{sedfit}

To obtain a first estimate of the physical parameters of the catalog sources, the observed SEDs
were fit using both \emph{i}) the grid of models from \citet{rob06} and the SED-fitting tool of
\citet{rob07}, based on the YSO/disk/envelope model of \citet{whi03}, and \emph{ii}) the simple
greybody model (see Eq.~\ref{greyb}). The \textsl{Stage 3b} source
catalog (Table~\ref{stats}) was used for the analysis: this reduced
the final investigated subsample to 318 objects for $\ell=30^{\circ}$, and~101 at
$\ell=59^{\circ}$, respectively.

In order to achieve a wider SED coverage, ancillary data from other Galactic plane surveys were
exploited to find possible counterparts and associate them according to the same criteria described
above. In particular, on MIPSGAL maps at 24~$\mu$m \citep{car09} a source extraction was
performed in the same way as for the Hi-GAL bands \citep{mol10b}, returning 294 counterparts
at $\ell=30^{\circ}$, and 89 at $\ell=59^{\circ}$, respectively. In addition, sources in the
1.1~mm band were retrieved from the Bolocam Galactic Plane Survey catalog \citep[BGPS,][]{ros09,agu10}.
A large difference in number is found between the millimeter counterparts available in the
$\ell=30^{\circ}$ and $\ell=59^{\circ}$ regions; the final number of BGPS counterparts
associated with \textsl{Stage 3b} sources in these two fields is 105 and only 4, respectively.
This discrepancy arises because the $\ell=59^{\circ}$ field is
relatively far from the inner Galaxy, compared with $\ell=30^{\circ}$, then a smaller number
of millimeter sources along the line of sight is reasonably expected.

\begin{figure}
   \centering
   \includegraphics[width=7.0cm, ]{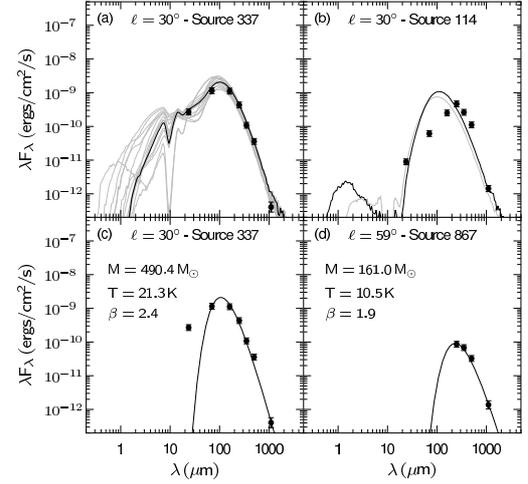}
   \caption{Panels (a) and (b): example of good and bad fit for two sources from $\ell=30^{\circ}$,
   respectively, obtained with the \citet{rob06} fitting procedure. The black filled circles with error
   bars show the MIPS and \emph{Herschel} data. The black solid line represents the best fit; the
   grey solid lines in panel (a) show all other models providing a good fit to the data. For the
   badly fit source only the best fit is shown.
Panel (c): the same SED of panel (a), but this time fitted with a greybody. Panel (d): the SED of a
source from $\ell=59^{\circ}$, with only SPIRE+BGPS data, again fitted with a greybody. Physical
parameters resulting from the greybody fit are also reported in panels (c) and (d), respectively.}
    \label{tom}
    \end{figure}

The model grid of \citet{rob06} covers a large range of stellar masses (from 0.1 to 50~$M_{\odot}$), and evolutionary stages (from embedded protostars to pre-main-sequence stars with low-mass circumstellar disks). The fitting tool uses linear regression to identify the best fitting models, allowing the interstellar extinction $A_V$ and distance to be free parameters within user-defined ranges. All \emph{Herschel} and MIPS 24~$\mu$m fluxes were assigned 20\% uncertainties, and all models that fit with a $\chi^2$ value satisfying $\chi^2 - \chi_{\rm best}^2 \le 3 \times n_{\rm data}$ were considered good fits. An interstellar extinction range of 0 to 20 was explored. Examples of a good and bad fits are shown in the top two panels of Fig.~\ref{tom}.
Given the high number of free parameters in the models and the relatively limited number of SED points we concentrated our attention on the macroscopic indications that can be provided by the model, namely the bolometric luminosity and the envelope mass. We took all values of these two parameters for the set of fits considered acceptable (see above) and the median values are estimated and used as $M_{\rm env}$ and $L_{\rm bol}$ in our subsequent analysis.


The SED-model fit provides acceptable results according to the above criteria for 127 source
in the $\ell=30^{\circ}$  field, and 43 sources in the $\ell=59^{\circ}$ field. For the remaining
sources no acceptable fits were found in the entire grid of models. A visual inspection showed that
they are mostly sources where the SED peaks at $\lambda \geq$170\um. These situations generally
correspond to cold envelopes with an important fraction of the gas at temperatures T$\leq$30K,
not considered in the model grid of \citet{rob06}. These SEDs were fitted with a greybody function
(Eq.~\ref{greyb}) which is probably more adequate to describe early-stage (or pre-stellar) cores than
the complex case of a protostar embedded in a dense envelope.
The free parameters derived from weighted least-squares fitting (corresponding to the
minimum $\chi^2$ value) are mass, $\beta$ index and temperature. The lower panels of Fig.~\ref{tom} show two examples
of SED best fit from $\ell=30^{\circ}$ and $\ell=59^{\circ}$, respectively. They are chosen to display two
different SED typologies: i) a source detected in all bands, with a 24~$\mu$m flux, which suggests
an embedded warm YSO, and ii) a source whose SED is composed only of SPIRE (250, 350 and 500 $\mu$m) and BGPS (1.1~mm)
fluxes that can be more reliably fitted with a greybody model.

As in the case shown in panel (c) of Fig.~\ref{tom}, in our sample the greybody fits generally fail
in reproducing fluxes at wavelengths $\leq 70~\mu$m, where the approach based on \citet{rob07} turns out to be
more appropriate.

From the fitted \textsl{Stage 3b} sample of sources, average temperatures and dust emissivity indices
were calculated. No noticeable differences emerge from the temperature distribution: the average
temperature and standard deviation are $\langle T_{30}\rangle=23.7~K$, $\sigma_{T_{30}}=13.8~K$, and
$\langle T_{59}\rangle=22.0~K$, $\sigma_{T_{59}}=8.7~K$ for the two fields, respectively. On the other
hand, the distribution of $\beta$ values appears to be more peaked for $\ell=59^{\circ}$ ($\langle\beta_{59}\rangle=1.0~K$,
$\sigma_{\beta_{59}}=0.8~K$) than for $\ell=30^{\circ}$ ($\langle\beta_{30}\rangle=1.6~K$, $\sigma_{\beta_{30}}=1.0~K$),
and centered on smaller values.


\section{Evolutionary timeline for massive YSOs}
\begin{figure}
   \centering
   \includegraphics[width=8.5cm]{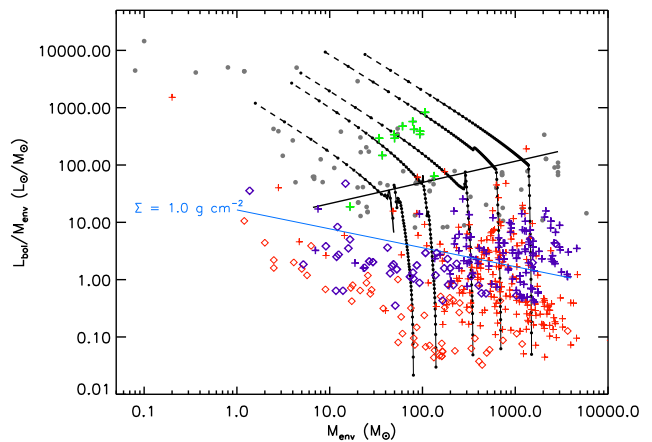}
   \caption{$L_{\rm bol}/M_{\rm env}$ vs $M_{\rm env}$ plot for fitted SEDs in the two
   investigated fields (crosses for $\ell=30^{\circ}$, and diamonds for $\ell=59^{\circ}$,
   respectively). The blue and red colors refer to fits obtained using the \citet{rob06}
   method (except the green points identifying models compatible with a central ZAMS star) and the greybody,
   respectively. The remaining grey filled circles represent the sources studied in
   \citet{mol08} (see their Fig.~9) and are plotted for comparison. Black lines represent
   the models (see text); dots mark $10^{4}$ yr intervals. The colored line represents the
   $L/M$ threshold from \citet{kru08}}.
    \label{lmdiag}
    \end{figure}

The physical parameters obtained from the SED fitting can be used to infer the evolutionary stage of the
observed sources by means of tools like the plot of the bolometric luminosity $L_{\rm bol}$ of a YSO and the
total envelope mass $M_{\rm env}$ as resulting from the fits. It has been exploited by \citet{sar96} and
\citet{mol08} to describe the evolutionary sequence for YSOs in the low mass and high mass regimes, respectively.
Indeed, sources in different stages are expected to occupy different regions of this diagram: objects dominated
by emission at large wavelengths ($\lambda \geq 250~\mu$m) should have a $L_{\rm bol}/M_{\rm env}$ ratio smaller
than that of more evolved ones. Here we prefer to plot the $L_{\rm bol}/M_{\rm env}$ ratio vs M$_{env}$
(Fig.~\ref{lmdiag}), also adapting predictions obtained from \citet{mol08} using the \citet{mck03} model of
collapse in turbulence-supported cores. This model describes the standard free-fall accretion of an envelope
onto a central core as a function of time, depicting evolutionary tracks that depend on the initial value
of $M_{\rm env}$ and on the final value of the central star mass $m_{\star}$ when it supposedly joins the ZAMS.
For the sake of brevity we refer the reader to the two papers mentioned above for a more exhaustive explanation; here we provide
a brief and qualitative description. Each track starts from its right-bottom end, and at first proceeds almost
vertically. In a region of the plane corresponding to the last third of the ascending tracks \citet{mol08} found sources
(open grey circles in Fig.~\ref{lmdiag}) whose SEDs were not consistent with an embedded ZAMS star, and were then
proposed as young and accreting pre-ZAMS protostars. The end of the accretion determines the end of the rapidly
ascending tracks. In this region of the plot, where sources similar to YSOs associated with UCHII regions are
found, their SED is compatible with embedded ZAMS. Starting from here, the tracks are essentially
determined by residual accretion and envelope dispersal due to molecular outflows and stellar winds. This region is
populated with ZAMS objects whose circumstellar envelope is of lesser and lesser importance.

Most of the points representing the sources of the present paper occupy a region Fig.~\ref{lmdiag} which
corresponds to the accretion phase part of the evolutionary tracks. Crosses and diamonds represent sources
for the $\ell=30^{\circ}$ and $\ell=59^{\circ}$ fields, respectively. In red are the objects which were
fitted with a greybody. In blue we plot the sources which are successfully fitted with \citet{rob06} models,
and where the parameters of the central embedded forming object \textit{are not} compatible with a ZAMS star
for more than 60\% of the models deemed acceptable. In green we plot the sources which could still
be fitted as above, but where this time the central object has the properties characteristic of ZAMS stars
in more than 40\% of the acceptable models.
It is reassuring that this first attempt at source classification confirms earlier results, in that objects
presently modeled as embedded ZAMS are indeed located where ZAMS are predicted to be found.


A quite clear distinction emerges in Fig.~\ref{lmdiag} between sources of $\ell=30^{\circ}$ and
$\ell=59^{\circ}$. On the one hand, an evident selection effect is present on luminosities
and masses for $\ell=30^{\circ}$ due to the higher typical distance (a factor 3 compared
with the sources in the $\ell=59^{\circ}$ field), on the other hand almost all the sources
from $\ell=59^{\circ}$ have luminosities below $\sim 10^3~L_{\odot}$, despite a wide range
of mass values. This would suggest a global difference from the evolutionary point of view
between the population detected in the two considered Hi-GAL fields, influenced by the
properties of the star-forming regions they host. For example, evidence of star formation
in a more advanced stage in the $\ell=30^{\circ}$ field are found in \citet{bal10}.

Figure~\ref{lmdiag} then suggests that most of the objects detected at present and for which the SED could be
reliably determined are in a very early (pre-ZAMS) phase of evolution. With the present data it is
not possible to ascertain if all detected cores are actively forming stars (and of which mass) or
if the population fraction at very low $L/M$ may be in a pre-protostellar phase. High $L/M$ ratios, however,
are difficult to explain without an actively forming high-mass star. To be more quantitative we
show in Fig.~\ref{lmdiag} the $L/M$ threshold corresponding to the critical core surface density
of 1~g~cm$^{-2}$ for the formation of massive stars as can be derived according to \citet{kru08}.

\begin{figure}
   \centering
   \includegraphics[width=7.5cm]{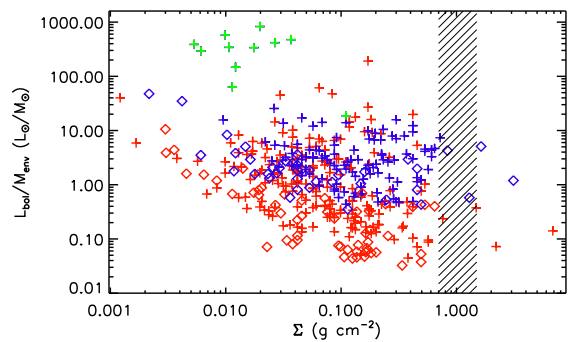}
   \caption{Plot of $L_{\rm bol}/M_{\rm env}$ vs core surface density in the two investigated fields (symbols are
   the same as in Fig.~\ref{lmdiag}, except for green crosses, which are not plotted here). The
   hashed area marks $\Sigma=1$~g~cm$^{-2}$.}
    \label{density}
    \end{figure}

Based on this threshold, the fraction of cores actively forming massive stars would amount to 47\% for $\ell=30^{\circ}$ and
13\% for $\ell=59^{\circ}$. However, after having calculated the surface density $\Sigma$ as the ratio between the core mass
and its area in the sky at 250~$\mu$m, we find it quite puzzling that the cores we model exhibit surface densities
well below the critical value. Figure~\ref{density} shows that the surface densities calculated for most of the
cores are lower than the critical values needed to form stars in the range between 10 and 200 $M_{\odot}$
(the shaded area in the figure). It can be argued that the adoption of different dust mass opacities
$\kappa_0$ could shift the distribution of points to the right, and partially beyond, the critical threshold.
It is, however, difficult to explain why sources with $L/M$ above the critical threshold
(above the sloping blue line in Fig.~\ref{lmdiag}) do not occupy a specific region of the plot,
but instead exhibit a scatter similar to the other sources in each region.

\begin{acknowledgements}
Data processing and maps production has been possible
thanks to ASI generous support via contract I/038/080/0.
\end{acknowledgements}

\end{document}